\documentclass[runningheads]{llncs}

\usepackage[T1]{fontenc}
\usepackage{graphicx}
\usepackage{amsmath}
\usepackage{amssymb}
\usepackage{booktabs}
\usepackage{multirow}
\usepackage{array}
\usepackage{xcolor}
\usepackage[hidelinks,colorlinks=true,linkcolor=black,citecolor=black,urlcolor=blue]{hyperref}
\usepackage{url}

\begin{document}

\title{BCER Agent: Reliable Long-Horizon MRI Workflow Execution via Compilation, Artifact Binding, and Bounded Local Recovery}

\titlerunning{BCER Agent}

\author{Ziyang Long\inst{1,2} \and
Xinqi Li\inst{2,3} \and
Junzhou Chen\inst{2} \and
Yifan Gao\inst{2,4} \and
Debiao Li\inst{2,1} \and
Hsin-Jung Yang\inst{2}}

\authorrunning{Z. Long et al.}

\institute{University of California, Los Angeles, Los Angeles, CA, USA \\
\email{ziyanglong@ucla.edu} \and
Biomedical Imaging Research Institute, Cedars-Sinai Medical Center, Los Angeles, CA, USA \\
\email{\{Junzhou.Chen,Yifan.Gao\}@cshs.org}; \email{\{Debiao.Li,hsin-jung.yang\}@csmc.edu} \and
Berlin Ultrahigh Field Facility, Max Delbr\"uck Center for Molecular Medicine in the Helmholtz Association, Berlin, Germany \\
\email{xinqi.li@mdc-berlin.de} \and
Tsinghua University, Beijing, China}

\maketitle

\begin{abstract}
Many recent medical VLM and agent studies are benchmarked on 2D images or comparatively short tool-calling exchanges, whereas real MRI analysis typically demands long, interdependent pipelines that operate on 3D/4D volumetric data. Under these conditions, reactive tool-calling agents are prone to cascading breakdowns triggered by faulty intermediate references, mismatched tool arguments, and limited control over cross-step dependencies. To address this, we introduce BCER (Brain--Cerebellum--Extremity--Reflector), a controller architecture aimed at dependable long-horizon MRI workflow execution. BCER decouples high-level planning from execution and provides bounded local recovery. We assess BCER on a multi-organ MRI benchmark covering brain, prostate, and cardiac tasks with both short- and long-chain workflows, using matched task contracts across controller variants and several backbone models. Relative to reactive baselines, BCER yields consistent improvements in end-to-end execution, with the most pronounced gains observed on long-chain workflows. BCER additionally enables auditability by maintaining explicit links between final outputs and intermediate artifacts and measurements. Code and benchmark are released at \url{https://github.com/Albertlongzi/BCER}.

\end{abstract}

\section{Introduction}
Recent work on medical VLMs and agents has reported encouraging progress on medical image understanding and tool-augmented reasoning~\cite{medagents,li2024mmedagent,kim2024mdagents,medagentpro}. Yet a large share of existing evaluations still focuses on relatively narrow settings, including pre-selected image inputs, slice- or case-level interpretation~\cite{li2023llavamed,nath2025vila}, visual question answering~\cite{lau2018vqarad,liu2021slake}, or short tool-use traces~\cite{nath2025vila}. In clinical practice, however, MRI analysis is more naturally cast as a multi-stage workflow rather than a one-hop question-answering exercise: it can start upstream (sometimes from reconstruction), traverse several dependent stages, and require consistent management of intermediate outputs across tools.

This raises two practical needs for deployable medical agents that remain insufficiently explored: (i) dependable execution over intermediate artifacts and cross-step dependencies, and (ii) auditability—linking conclusions back to intermediate evidence and measurements~\cite{rajpurkar2022aihealth}.

Many current tool-augmented baselines and agent frameworks rely on reactive, interleaved planning-and-acting loops (e.g., ReAct-style prompting~\cite{yao2022react,shinn2024reflexion,medagents}). Although effective for flexible short-form interactions, such designs offer limited explicit control over intermediate artifact dependencies~\cite{wang2024codeact}, an aspect that becomes increasingly critical for long MRI workflows~\cite{gorgolewski2011nipype}.

In this study, we examine an execution gap that arises in long-horizon MRI agent workflows: failures stemming from incorrect intermediate-result references, incompatible tool arguments, or broken step dependencies can accumulate and prevent task completion, even when the underlying tools are accurate. Accordingly, we elevate execution reliability—covering both end-to-end completion and robust handling of execution failures—to a first-class objective.

To this end, we present BCER (Brain--Cerebellum--Extremity--Reflector), a controller architecture designed for reliable long-horizon MRI workflow execution. BCER separates high-level planning from constrained execution and bounded local recovery, supporting end-to-end workflows whose intermediate artifacts remain traceable for downstream review.

Our main contributions are three-fold:
\begin{itemize}
\item \textbf{A workflow-oriented MRI agent setting for multi-organ research pipelines.} We formulate medical agents around multi-step MRI workflows beyond diagnostic QA, targeting research-oriented pipelines spanning organs and tasks, including workflows that may originate from raw data when available.
\item \textbf{BCER: plan--execution separation with bounded local recovery.} We propose the Brain--Cerebellum--Extremity--Reflector (BCER) architecture, which decouples goal-level planning from constrained execution and enables localized repair-and-retry instead of full workflow restarts, improving robustness in long MRI tool chains while keeping intermediate artifacts traceable.
\item \textbf{A task-contract benchmark for short- and long-chain MRI workflows.} We introduce a task-contract-based evaluation that jointly measures task success and completion fidelity across controller variants and backbone models.
\end{itemize}

\begin{figure}[t]
\centering
\includegraphics[width=0.95\linewidth]{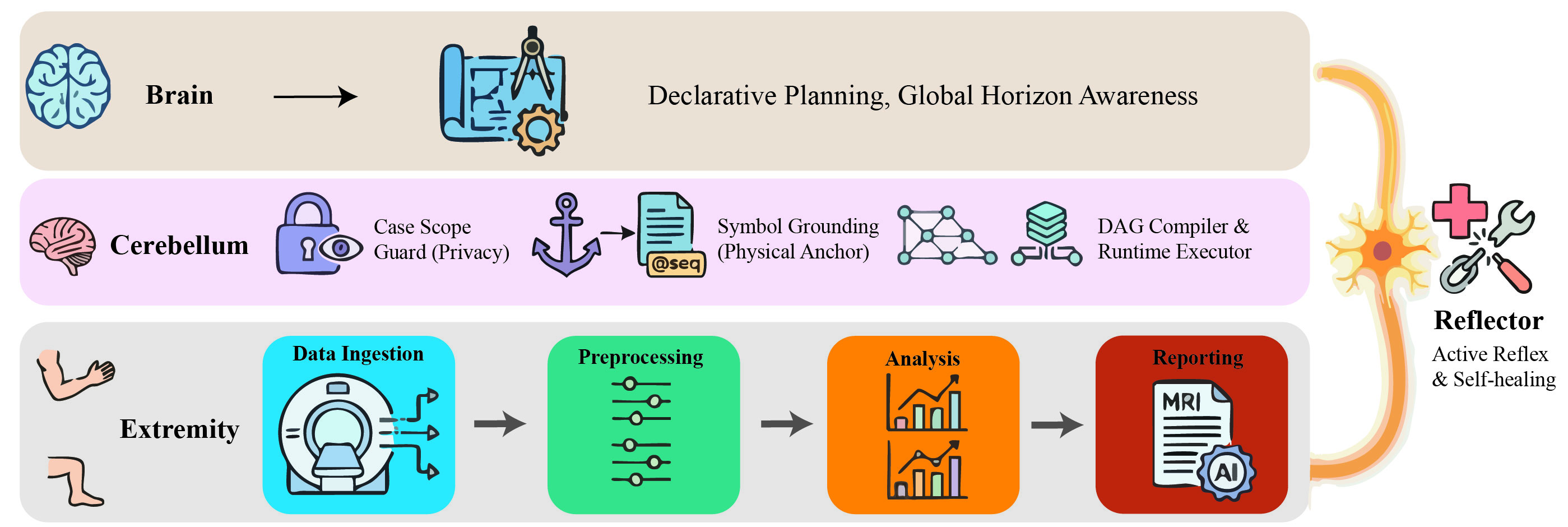}
\caption{BCER overview. The Brain produces a constrained plan sketch from the user goal and the available MRI inputs. The Cerebellum compiles the sketch into an executable workflow graph and runs it under run-time constraints, dispatching tools from the Extremity (MRI tool library) while logging intermediate artifacts and outcomes. The Reflector carries out bounded step- or sub-workflow repair whenever failures occur.}
\label{fig:overview}
\end{figure}

\section{Method}

\subsection{Problem Setup}
We consider an MRI agent task in which the system is provided with volumetric imaging data $V$, accessible metadata $M$, and a natural-language user goal $g$:
\begin{equation}
I = (V, M, g).
\end{equation}

The objective is not only to return a final answer, but also to complete a multi-step MRI workflow and deliver the required task outputs (e.g., processed volumes, masks, measurements, classifications, or a structured report), along with execution records that support traceability.

We define success via task-contract satisfaction: every required workflow step and output specified by the task must be produced under the system's execution constraints. This formulation enables evaluation of both end-to-end task completion and step-level completion within tool chains.

\subsection{BCER}

\textbf{Brain.} The Brain produces a plan sketch from $(V, M, g)$ that outlines intended workflow steps and constraints but is not directly executable. This split between conceptual planning and low-level execution borrows from prior modular agent systems~\cite{wang2023describe,shen2023hugginggpt}, adapted here to the demands of MRI workflows. Specifically, the Brain neither executes tools nor manipulates file paths directly, and it may leave some executable fields to be filled in further downstream.

\textbf{Extremity.} The Extremity serves as BCER's executable tool layer. It exposes standardized MRI-facing tools for data I/O, preprocessing, analysis, quantification, and reporting, with uniform argument schemas, typed outputs, and normalized error codes for runtime dispatch and recovery. The library combines wrapped conventional operators (e.g., reconstruction, denoising, registration) with integrated task-specific modules drawn from established MRI subproblems (e.g., segmentation, detection, and correction)~\cite{adams2022prostate158,saha2021end,long2026let,zhao2025reverse,myronenko20183d}, along with workflow utilities for intermediate-state resolution (such as ED/ES frame selection) and evidence-backed decision tools that fuse measurements with explicit rules/skills for downstream classification and report generation.

\textbf{Cerebellum.} The Cerebellum is the runtime controller that turns the Brain's high-level plan sketch into an executable workflow. The sketch is not run as-is because it may be only partially specified and may omit dispatch-level details. Instead, the Cerebellum checks the proposed steps against task-level interface and dependency constraints, fills in only those defaults or unambiguous links that dispatch requires, and yields a runnable workflow graph. In this sense, compilation does not instantiate a fixed task template; rather, it transforms a partially specified plan into an executable form with explicit dependencies and bounded execution semantics.

Prior to each tool call, the Cerebellum resolves references to inputs and intermediate results through symbolic tokens: \texttt{@seq.*}, \texttt{@node.<id>.<field>}, \texttt{@case.*}, and \texttt{@runtime.*}. These tokens are bound to concrete values/paths within the current case run so that tools consume real artifacts instead of LLM-generated strings. Inspired by artifact-passing mechanisms in modular tool-use agents~\cite{shen2023hugginggpt}, this explicit symbolic binding provides a more robust dataflow interface for large medical artifacts. The binding is a key differentiator from direct ReAct-style tool calling and is isolated empirically through the ReAct+Bind controller variant.

The Cerebellum runs nodes once their prerequisites are met, updates per-case state along with an event trace recording statuses/outputs/errors, and enforces run-time constraints (e.g., case-level scope/sandbox boundaries). As a result, failures surface as explicit node failures with traceable context, instead of silently corrupting downstream steps.

\textbf{Reflector.} The Reflector deals with failures of required nodes at runtime through bounded local repair. Earlier reflection-style mechanisms in LLM agents typically operate at the granularity of full-turn or wholesale response revision~\cite{shinn2023reflexion,madaan2023selfrefine}, which becomes increasingly expensive within long MRI workflows. Rather than restarting the workflow, the Reflector inspects the failed node, its arguments, the error message, and available artifacts, and proposes a minimal retry patch (e.g., fixing a token reference, removing an invalid override, or applying a safe argument adjustment). The Cerebellum then retries only the failed node (or a small affected sub-workflow) under the same run-time constraints, while preserving successful upstream results. Repair follows a two-stage strategy: deterministic fixes are tried first, and a constrained LLM-based repair is invoked only if needed. This improves robustness in long tool chains while keeping recovery localized and auditable.

\begin{figure}[t]
\centering
\includegraphics[width=0.98\linewidth]{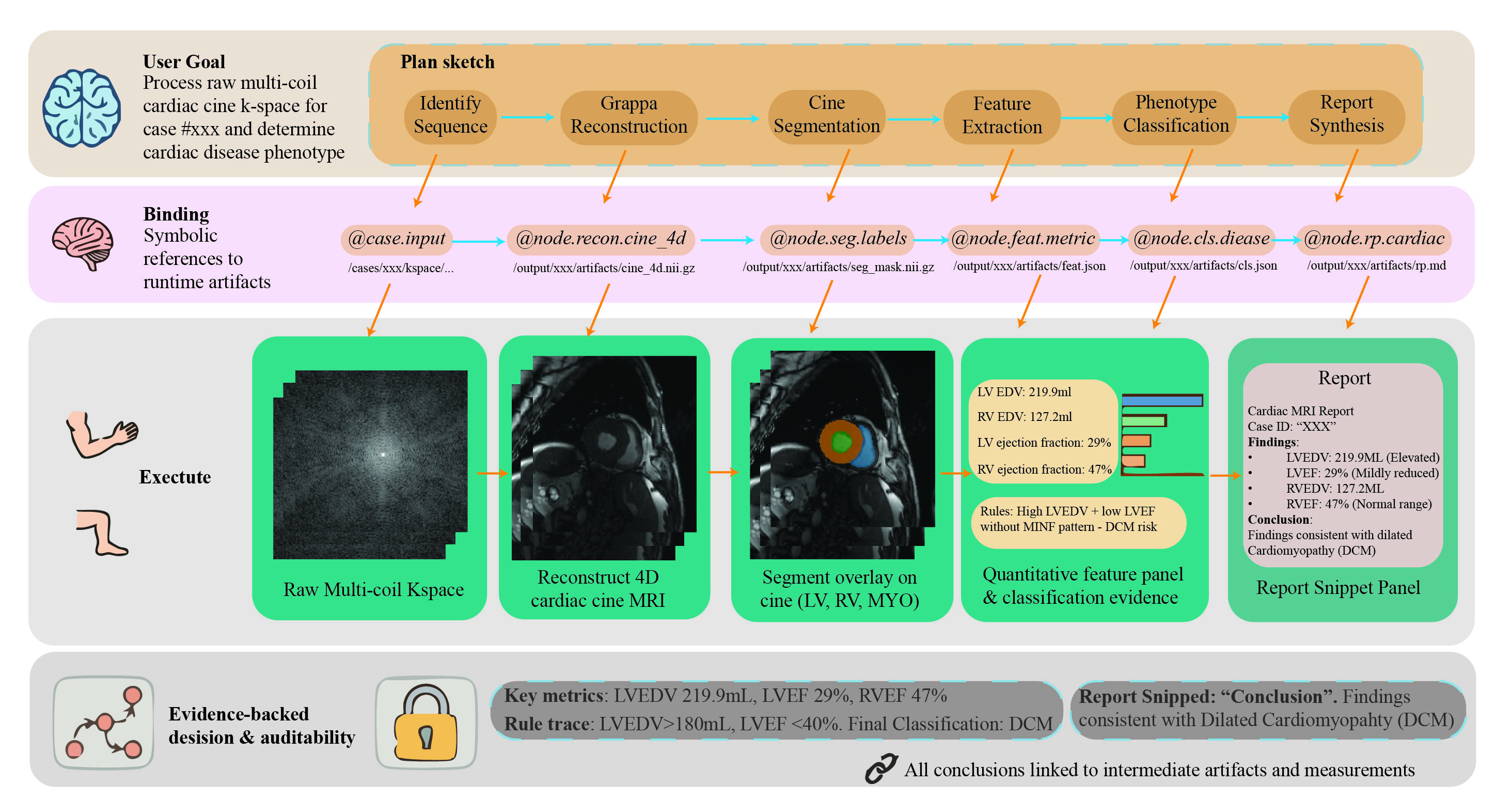}
\caption{BCER end-to-end cardiac workflow walkthrough. From top to bottom, the figure depicts four linked layers of execution. Top: a plan sketch derived from the user goal, here shown as a multi-step cardiac pipeline that proceeds from sequence identification and reconstruction through segmentation, feature extraction, phenotype classification, and report synthesis. Second row: symbolic artifact binding, where abstract references (e.g., \texttt{@case.input}, \texttt{@node.*}) are resolved to concrete runtime outputs. Third row: execution outputs produced by the tool chain, including reconstructed cine MRI, segmentation overlays, quantitative measurements, and the report snippet. Bottom: evidence-backed traceability, where final conclusions stay linked to intermediate artifacts, measurements, and rule traces for downstream review.}
\label{fig:walkthrough}
\end{figure}

BCER constitutes a domain-structured integration of planning, compilation, constrained execution, and local recovery for MRI workflows, rather than a claim that each component is independently novel. In contrast to open-ended symbolic planning or directly executable LLM plans, the Brain emits a constrained sketch that the Cerebellum compiles into a typed, case-scoped workflow graph with dependency resolution and symbolic artifact binding. Figure~\ref{fig:walkthrough} provides an end-to-end cardiac walkthrough that illustrates how BCER converts a user goal into a compiled multi-step workflow with staged execution outputs and evidence-linked reporting.

\section{Experiments and Results}

\subsection{Benchmark Setup}
We rely on four public MRI datasets spanning brain, prostate, and cardiac domains: BraTS 2018~\cite{bakas2018identifying}, fastMRI Prostate~\cite{tibrewala2024fastmri}, ACDC~\cite{bernard2018deep}, and CMRxRecon~\cite{wang2024cmrxrecon}. The benchmark comprises 8 tasks covering denoising, super-resolution, segmentation, reconstruction, registration, grading/classification, and end-to-end report generation (Table~\ref{tab:tasks}). Each run is parameterized by a task contract that specifies the input case, the goal, the allowed tool subset, and the required outputs (gold deliverables). The same contracts are used across controller arms, producing matched action spaces and consistent success criteria.

\begin{table}[t]
\centering
\caption{Benchmark task suite overview. Tasks are defined by task contracts and cover both short-chain (S) and long-chain (L) MRI workflows across brain, prostate, and cardiac domains. The table lists the input type, organ domain, task-specific case count, and the required gold output for each task.}
\label{tab:tasks}
\setlength{\tabcolsep}{4pt}
\begin{tabular}{lcllrl}
\toprule
Task & Chain & Input & Organ & Cases & Gold Output \\
\midrule
Denoise     & S & NIfTI            & brain / prostate & 200 & denoised volume \\
SuperRes    & S & NIfTI            & multi-organ      & 310 & super-resolved volume \\
Segment     & S & DICOM / NIfTI    & brain            & 100 & organ mask \\
Recon       & S & k-space          & cardiac          & 10  & reconstructed image volume \\
Register    & S & DICOM / NIfTI    & prostate         & 100 & aligned volume \\
BrainGrade  & L & NIfTI            & brain            & 100 & grade/class label \\
ProstateRpt & L & DICOM            & prostate         & 100 & structured report \\
CardiacRpt  & L & k-space / DICOM  & cardiac          & 110 & structured report \\
\midrule
Total       & -- & --              & --               & 1030 & -- \\
\bottomrule
\end{tabular}
\end{table}

We group tasks into short-chain (single- or few-step imaging operations) and long-chain (multi-stage analysis and reporting workflows), and use this split throughout the results. We evaluate four controller variants in end-to-end execution: \textbf{ReAct}, \textbf{ReAct+Bind}, \textbf{ReAct+Bind+Ref}, and \textbf{BCER}. These variants form a progressive ablation ladder: starting from direct reactive tool calling, then adding symbolic artifact binding, then local repair, and finally BCER's full plan--execution control stack. This setup enables both comparison against a standard reactive baseline and attribution of gains to each added controller component. We first compare these controllers on the benchmark tasks under normal execution conditions, and then test whether the same trends persist across multiple backbone models.

As metrics, we report \textbf{Success Rate (SR)} and \textbf{Task Completion Rate (TCR)} against fixed task contracts that specify inputs, goals, allowed tools, required milestones, and deliverables. SR is binary per case: all required milestones and deliverables must be produced and validated. TCR is the fraction of reference contract milestones that are validated, rather than attempted or generated steps; arbitrary files or free-form answers do not count. These metrics measure controller-level execution reliability, not clinical diagnostic accuracy or standalone tool performance.

\subsection{End-to-end controller comparison}
Table~\ref{tab:main} compares task-level SR/TCR for the four controller variants using Qwen3-VL-30B-A3B-Thinking as the backbone. The dominant pattern is not simply that performance climbs from ReAct to BCER, but that the principal failure mode also shifts at each stage of the controller stack.

\begin{table}[t]
\centering
\caption{Task-level success rate (SR, \%) / task completion rate (TCR, \%) for the end-to-end controller comparison under normal execution. Tasks are grouped into short-chain and long-chain workflows.}
\label{tab:main}
\setlength{\tabcolsep}{6pt}
\begin{tabular}{lcccc}
\toprule
Task & ReAct & ReAct+Bind & ReAct+Bind+Ref & BCER \\
\midrule
\multicolumn{5}{l}{\textit{Short-chain tasks}} \\
Denoise           & 95/95   & 98/98   & 100/100 & 100/100 \\
Super-resolution  & 75/75   & 90/90   & 95/95   & 100/100 \\
Segmentation      & 89/91   & 98/99   & 100/100 & 100/100 \\
Reconstruction    & 100/100 & 100/100 & 100/100 & 100/100 \\
Registration      & 69/77   & 100/100 & 100/100 & 100/100 \\
\midrule
\multicolumn{5}{l}{\textit{Long-chain tasks}} \\
Brain grading     & 70/82   & 100/100 & 100/100 & 100/100 \\
Prostate report   & 0/46    & 0/72    & 0/72    & 99/99 \\
Cardiac report    & 0/22    & 63/66   & 88/89   & 93/93 \\
\midrule
Total             & 65/74   & 83/90   & 87/95   & 99/99 \\
\bottomrule
\end{tabular}
\end{table}

Under ReAct, failures are largely driven by unstable step execution, including parameter/schema mismatches as well as path or input-selection errors. On long-chain tasks, the issue is often not a single missing step, but rather a single unstable tool call that breaks downstream dependencies and collapses the entire workflow.

Introducing binding (\emph{ReAct+Bind}) substantially reduces low-level intermediate-reference and path-resolution errors, which accounts for the sizeable gain in overall SR/TCR (from 65/74 to 83/90). However, many remaining failures are no longer simple reference-grounding mistakes; instead, they reflect workflow-level omissions, such as missing a required tool step or producing an incomplete processing chain.

Adding the Reflector (\emph{ReAct+Bind+Ref}) further suppresses low-level execution errors through local repair, lifting overall performance to 87/95. Even so, it does not fully resolve failures caused by incomplete workflow structure. In our error review, the remaining cases are dominated by missing or insufficiently specified steps, together with occasional task-contract misses (e.g., incomplete completion of required resampling in a small number of short super-resolution cases).

By contrast, BCER removes most of these structural failure modes earlier, at the planning-and-compilation stage. Relative to ReAct+Bind+Ref, BCER produces more complete workflow graphs and more consistent step dependencies, so failures are less likely to originate from hallucinated parameters, invalid paths, or omitted nodes. Consequently, the remaining errors cluster around the robustness limits of a small number of tools or task contracts, rather than cascading controller failures. This contrast is most apparent on long-chain reporting tasks: whereas ReAct fails almost completely on ProstateRpt and CardiacRpt (0/46 and 0/22), BCER reaches 99/99 and 93/93, indicating that compiled execution improves not only reference grounding but also overall workflow integrity.

\subsection{Across Backbone Models}

\begin{figure}[t]
\centering
\includegraphics[width=0.95\linewidth]{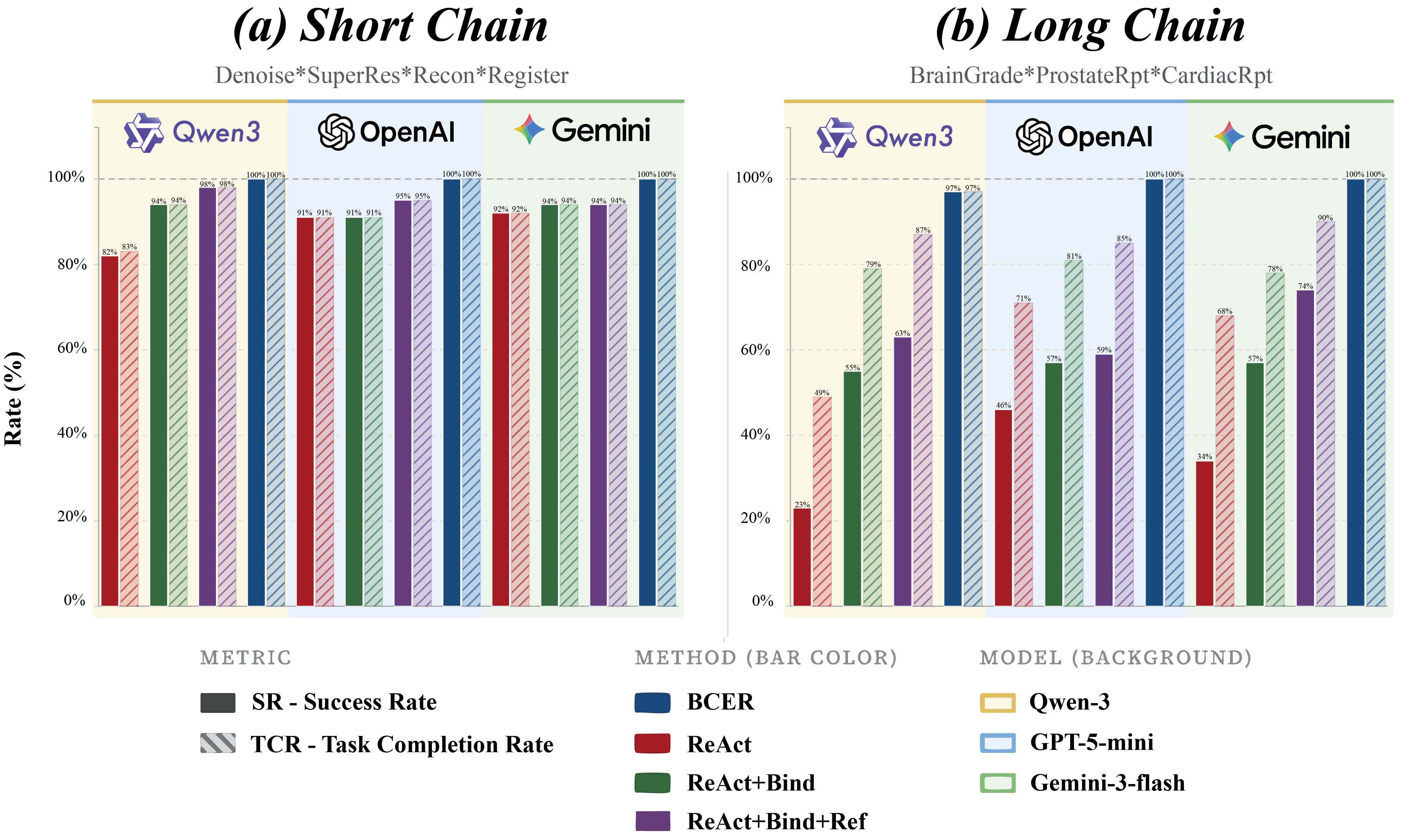}
\caption{Bars report SR (solid) and TCR (hatched) for ReAct, ReAct+Bind, ReAct+Bind+Ref, and BCER, aggregated over short-chain tasks (Denoise, SuperRes, Recon, Register) and long-chain tasks (BrainGrade, ProstateRpt, CardiacRpt).}
\label{fig:backbones}
\end{figure}

Figure~\ref{fig:backbones} extends the controller comparison across multiple LLM backbones. On short-chain tasks, all backbones perform relatively well and the gap between methods is modest. On long-chain tasks, however, the gap widens substantially: plain ReAct degrades across all three backbones, and stronger base models only partially mitigate this drop. In contrast, adding Bind and then Reflector yields consistent gains for every backbone, suggesting that the principal bottleneck is not solely model capacity but also execution control over intermediate artifacts and workflow dependencies. Across all evaluated backbones, BCER remains the most stable variant and approaches saturation on long chains, indicating that the benefits of compiled execution generalize beyond any single base model.

\section{Conclusion}
We have introduced BCER, a workflow controller architecture for long, multi-step MRI analysis that combines plan--execution separation, symbolic artifact binding, and bounded local recovery. Across a multi-organ MRI benchmark with both short- and long-chain tasks, BCER consistently outperforms reactive tool-calling baselines, with the largest gains observed on long-chain workflows.

This work targets controller capability rather than the upper-bound performance of individual tools or backbone models. Under matched task contracts and tool availability, BCER improves executability by enforcing a stronger execution structure, at the cost of reduced flexibility. The present study is not a prospective clinical validation and does not claim to replace expert-generated reports or clinician judgment; expert knowledge enters through task definitions, public benchmark annotations where available, predefined deliverables, and rule/tool interfaces. The current design also assumes a curated tool library with explicit interfaces, rather than open-ended tool discovery. Future work will examine safer tool expansion, clinician-in-the-loop validation, and broader evaluation within prospective clinical research workflows.

Taken together, BCER is a practical step toward more reliable, evidence-linked medical imaging agents: it strengthens end-to-end workflow robustness while preserving traceability from final outputs back to intermediate artifacts and measurements, which is important for reproducibility, auditability, and the eventual translational use of MRI-centered clinical research.

\subsubsection*{Acknowledgments.} This work was supported in part by the National Institutes of Health under grants 1R01HL181091 and 5R01HL165211.

\clearpage
\bibliographystyle{splncs04}
\bibliography{references}

@misc{medagentpro,
  title={MedAgent-Pro: Towards Evidence-based Multi-modal Medical Diagnosis via Reasoning Agentic Workflow},
  author={Wang, Ziyue and Wu, Junde and Cai, Linghan and Low, Chang Han and Yang, Xihong and Li, Qiaxuan and Jin, Yueming},
  year={2025},
  eprint={2503.18968},
  archivePrefix={arXiv},
  primaryClass={cs.AI},
  doi={10.48550/arXiv.2503.18968},
  url={https://arxiv.org/abs/2503.18968}
}

@article{lau2018vqarad,
  title={A dataset of clinically generated visual questions and answers about radiology images},
  author={Lau, Jason J. and Gayen, Soumya and Ben Abacha, Asma and Demner-Fushman, Dina},
  journal={Scientific Data},
  volume={5},
  pages={180251},
  year={2018},
  doi={10.1038/sdata.2018.251},
  publisher={Nature Publishing Group}
}

@article{rajpurkar2022aihealth,
  title={AI in health and medicine},
  author={Rajpurkar, Pranav and Chen, Emma and Banerjee, Oishi and Topol, Eric J.},
  journal={Nature Medicine},
  volume={28},
  number={1},
  pages={31--38},
  year={2022},
  doi={10.1038/s41591-021-01614-0},
  publisher={Nature Publishing Group}
}

@article{gorgolewski2011nipype,
  title={Nipype: A Flexible, Lightweight and Extensible Neuroimaging Data Processing Framework in Python},
  author={Gorgolewski, Krzysztof and Burns, Christopher D. and Madison, Cindee and Clark, Dav and Halchenko, Yaroslav O. and Waskom, Michael L. and Ghosh, Satrajit S.},
  journal={Frontiers in Neuroinformatics},
  volume={5},
  pages={13},
  year={2011},
  doi={10.3389/fninf.2011.00013},
  publisher={Frontiers Media SA}
}

@inproceedings{wang2024codeact,
  title={Executable Code Actions Elicit Better LLM Agents},
  author={Wang, Xingyao and Chen, Yangyi and Yuan, Lifan and Zhang, Yizhe and Li, Yunzhu and Peng, Hao and Ji, Heng},
  booktitle={Proceedings of the 41st International Conference on Machine Learning},
  series={Proceedings of Machine Learning Research},
  volume={235},
  pages={50208--50232},
  year={2024},
  eprint={2402.01030},
  archivePrefix={arXiv},
  primaryClass={cs.CL},
  doi={10.48550/arXiv.2402.01030},
  url={https://proceedings.mlr.press/v235/wang24h.html}
}

@misc{wang2023describe,
  title={Describe, Explain, Plan and Select: Interactive Planning with Large Language Models Enables Open-World Multi-Task Agents},
  author={Wang, Zihao and Cai, Shaofei and Liu, Anji and Ma, Xiaojian and Liang, Yitao},
  year={2023},
  eprint={2302.01560},
  archivePrefix={arXiv},
  primaryClass={cs.CL},
  doi={10.48550/arXiv.2302.01560},
  url={https://arxiv.org/abs/2302.01560}
}

@misc{shen2023hugginggpt,
  title={HuggingGPT: Solving AI Tasks with ChatGPT and its Friends in Hugging Face},
  author={Shen, Yongliang and Song, Kaitao and Tan, Xu and Li, Dongsheng and Lu, Weiming and Zhuang, Yueting},
  year={2023},
  eprint={2303.17580},
  archivePrefix={arXiv},
  primaryClass={cs.CL},
  doi={10.48550/arXiv.2303.17580},
  url={https://arxiv.org/abs/2303.17580}
}

@misc{shinn2023reflexion,
  title={Reflexion: Language Agents with Verbal Reinforcement Learning},
  author={Shinn, Noah and Labash, Brian and Gopinath, Anurag and Narasimhan, Karthik and Yao, Shunyu},
  year={2023},
  eprint={2303.11366},
  archivePrefix={arXiv},
  primaryClass={cs.CL},
  doi={10.48550/arXiv.2303.11366},
  url={https://arxiv.org/abs/2303.11366}
}

@misc{madaan2023selfrefine,
  title={Self-Refine: Iterative Refinement with Self-Feedback},
  author={Madaan, Aman and Tandon, Niket and Gupta, Prakhar and Hallinan, Skyler and Gao, Luyu and Wiegreffe, Sarah and Alon, Uri and Dziri, Nouha and Qian, Yiming and Choi, Yejin},
  year={2023},
  eprint={2303.17651},
  archivePrefix={arXiv},
  primaryClass={cs.CL},
  doi={10.48550/arXiv.2303.17651},
  url={https://arxiv.org/abs/2303.17651}
}

@misc{liu2021slake,
  title={SLAKE: A Semantically-Labeled Knowledge-Enhanced Dataset for Medical Visual Question Answering},
  author={Liu, Bo and Zhan, Li-Ming and Xu, Li and Ma, Lin and Yang, Yan and Wu, Xiao-Ming},
  year={2021},
  eprint={2102.09542},
  archivePrefix={arXiv},
  primaryClass={cs.CV},
  doi={10.48550/arXiv.2102.09542},
  url={https://arxiv.org/abs/2102.09542}
}

@misc{li2023llavamed,
  title={LLaVA-Med: Training a Large Language-and-Vision Assistant for Biomedicine in One Day},
  author={Li, Chunyuan and Wong, Cliff and Zhang, Sheng and Usuyama, Naoto and Liu, Haotian and Yang, Jianwei and Naumann, Tristan and Poon, Hoifung and Gao, Jianfeng},
  year={2023},
  eprint={2306.00890},
  archivePrefix={arXiv},
  primaryClass={cs.CV},
  doi={10.48550/arXiv.2306.00890},
  url={https://arxiv.org/abs/2306.00890}
}

@inproceedings{medagents,
  author = {Tang, Xu and Zou, Andi and Zhang, Zheng and Li, Zhiyuan and Zhao, Yunyao and Zhang, Xiyang and Cohan, Arman and Gerstein, Mark},
  title = {MedAgents: Large Language Models as Collaborators for Zero-shot Medical Reasoning},
  booktitle = {Findings of the Association for Computational Linguistics: ACL 2024},
  year = {2024},
  pages = {599--621},
  address = {Bangkok, Thailand},
  publisher = {Association for Computational Linguistics},
  doi = {10.18653/v1/2024.findings-acl.33}
}

@article{kim2024mdagents,
  title={Mdagents: An adaptive collaboration of llms for medical decision-making},
  author={Kim, Yubin and Park, Chanwoo and Jeong, Hyewon and Chan, Yik S and Xu, Xuhai and McDuff, Daniel and Lee, Hyeonhoon and Ghassemi, Marzyeh and Breazeal, Cynthia and Park, Hae W},
  journal={Advances in Neural Information Processing Systems},
  volume={37},
  pages={79410--79452},
  year={2024}
}

@inproceedings{li2024mmedagent,
  title={Mmedagent: Learning to use medical tools with multi-modal agent},
  author={Li, Binxu and Yan, Tiankai and Pan, Yuanting and Luo, Jie and Ji, Ruiyang and Ding, Jiayuan and Xu, Zhe and Liu, Shilong and Dong, Haoyu and Lin, Zihao and others},
  booktitle={Findings of the Association for Computational Linguistics: EMNLP 2024},
  pages={8745--8760},
  year={2024}
}

@inproceedings{nath2025vila,
  title={Vila-m3: Enhancing vision-language models with medical expert knowledge},
  author={Nath, Vishwesh and Li, Wenqi and Yang, Dong and Myronenko, Andriy and Zheng, Mingxin and Lu, Yao and Liu, Zhijian and Yin, Hongxu and Law, Yee Man and Tang, Yucheng and others},
  booktitle={Proceedings of the Computer Vision and Pattern Recognition Conference},
  pages={14788--14798},
  year={2025}
}

@inproceedings{yao2022react,
  title={React: Synergizing reasoning and acting in language models},
  author={Yao, Shunyu and Zhao, Jeffrey and Yu, Dian and Du, Nan and Shafran, Izhak and Narasimhan, Karthik R and Cao, Yuan},
  booktitle={The eleventh international conference on learning representations},
  year={2022}
}

@article{shinn2024reflexion,
  title={Reflexion: Language agents with verbal reinforcement learning, 2023},
  author={Shinn, Noah and Cassano, Federico and Berman, Edward and Gopinath, Ashwin and Narasimhan, Karthik and Yao, Shunyu},
  journal={URL https://arxiv. org/abs/2303.11366},
  volume={8},
  year={2024}
}

@article{bernard2018deep,
  title={Deep learning techniques for automatic MRI cardiac multi-structures segmentation and diagnosis: is the problem solved?},
  author={Bernard, Olivier and Lalande, Alain and Zotti, Clement and Cervenansky, Frederick and Yang, Xin and Heng, Pheng-Ann and Cetin, Irem and Lekadir, Karim and Camara, Oscar and Ballester, Miguel Angel Gonzalez and others},
  journal={IEEE transactions on medical imaging},
  volume={37},
  number={11},
  pages={2514--2525},
  year={2018},
  publisher={ieee}
}

@article{wang2024cmrxrecon,
  title={CMRxRecon: A publicly available k-space dataset and benchmark to advance deep learning for cardiac MRI},
  author={Wang, Chengyan and Lyu, Jun and Wang, Shuo and Qin, Chen and Guo, Kunyuan and Zhang, Xinyu and Yu, Xiaotong and Li, Yan and Wang, Fanwen and Jin, Jianhua and others},
  journal={Scientific Data},
  volume={11},
  number={1},
  pages={687},
  year={2024},
  publisher={Nature Publishing Group UK London}
}

@article{bakas2018identifying,
  title={Identifying the best machine learning algorithms for brain tumor segmentation, progression assessment, and overall survival prediction in the BRATS challenge},
  author={Bakas, Spyridon and Reyes, Mauricio and Jakab, Andras and Bauer, Stefan and Rempfler, Markus and Crimi, Alessandro and Shinohara, Russell Takeshi and Berger, Christoph and Ha, Sung Min and Rozycki, Martin and others},
  journal={arXiv preprint arXiv:1811.02629},
  year={2018}
}

@article{tibrewala2024fastmri,
  title={FastMRI Prostate: A public, biparametric MRI dataset to advance machine learning for prostate cancer imaging},
  author={Tibrewala, Radhika and Dutt, Tarun and Tong, Angela and Ginocchio, Luke and Lattanzi, Riccardo and Keerthivasan, Mahesh B and Baete, Steven H and Chopra, Sumit and Lui, Yvonne W and Sodickson, Daniel K and others},
  journal={Scientific data},
  volume={11},
  number={1},
  pages={404},
  year={2024},
  publisher={Nature Publishing Group UK London}
}

@inproceedings{myronenko20183d,
  title={3D MRI brain tumor segmentation using autoencoder regularization},
  author={Myronenko, Andriy},
  booktitle={International MICCAI brainlesion workshop},
  pages={311--320},
  year={2018},
  organization={Springer}
}

@article{adams2022prostate158,
  title={Prostate158-An expert-annotated 3T MRI dataset and algorithm for prostate cancer detection},
  author={Adams, Lisa C and Makowski, Marcus R and Engel, G{\"u}nther and Rattunde, Maximilian and Busch, Felix and Asbach, Patrick and Niehues, Stefan M and Vinayahalingam, Shankeeth and van Ginneken, Bram and Litjens, Geert and others},
  journal={Computers in biology and medicine},
  volume={148},
  pages={105817},
  year={2022},
  publisher={Elsevier}
}

@article{saha2021end,
  title={End-to-end prostate cancer detection in bpMRI via 3D CNNs: Effects of attention mechanisms, clinical priori and decoupled false positive reduction},
  author={Saha, Anindo and Hosseinzadeh, Matin and Huisman, Henkjan},
  journal={Medical image analysis},
  volume={73},
  pages={102155},
  year={2021},
  publisher={Elsevier}
}

@article{long2026let,
  title={Let Distortion Guide Restoration (DGR): A physics-informed learning framework for Prostate Diffusion MRI},
  author={Long, Ziyang and Nader, Binesh and Wang, Lixia and Malaji, Archana Vadiraj and Yang, Chia-Chi and Sun, Haoran and Saouaf, Rola and Daskivich, Timothy and Kim, Hyung and Xie, Yibin and others},
  journal={arXiv preprint arXiv:2601.00226},
  year={2026}
}

@inproceedings{zhao2025reverse,
  title={Reverse Imaging for Wide-Spectrum Generalization of Cardiac MRI Segmentation},
  author={Zhao, Yidong and Kellman, Peter and Xue, Hui and Yang, Tongyun and Zhang, Yi and Han, Yuchi and Simonetti, Orlando and Tao, Qian},
  booktitle={International Conference on Medical Image Computing and Computer-Assisted Intervention},
  pages={555--565},
  year={2025},
  organization={Springer}
}

\end{document}